\documentclass{emulateapj}
\usepackage{multirow}
\usepackage{natbib}     
\usepackage{amsmath}

\usepackage{hyperref}
\usepackage[UKenglish]{babel}

\begin{document}

\title{AGN Heating in Simulated Cool-Core Clusters}

\author{Yuan Li\altaffilmark{1}, Mateusz Ruszkowski\altaffilmark{1}, Greg L. Bryan\altaffilmark{2, 3}}

\altaffiltext{1}{Department of Astronomy, University of Michigan, 1085 S University Ave, Ann Arbor, MI 48109; email: yuanlium@umich.edu}
\altaffiltext{2}{Department of Astronomy, Columbia University, Pupin Physics Laboratories, New York, NY 10027}
\altaffiltext{3}{Center for Computational Astronomy (CCA), New York, NY}

\begin{abstract}

We analyze heating and cooling processes in an idealized simulation of a cool-core cluster, where momentum-driven AGN feedback balances radiative cooling in a time-averaged sense. We find that, on average, energy dissipation via shock waves is almost an order of magnitude higher than via turbulence. Most of the shock waves in the simulation are very weak shocks with Mach numbers smaller than 1.5, but the stronger shocks, although rare, dissipate energy more effectively. We find that shock dissipation is a steep function of radius, with most of the energy dissipated within 30 kpc, while radiative cooling loses area less concentrated. However, adiabatic processes and mixing (of post-shock materials and the surrounding gas) are able to redistribute the heat throughout the core. A considerable fraction of the AGN energy also escapes the core region. The cluster goes through cycles of AGN outbursts accompanied by periods of enhanced precipitation and star formation, over Gyr timescales. The cluster core is under-heated at the end of each cycle, but over-heated at the peak of the AGN outburst. During the heating-dominant phase, turbulent dissipation alone is often able to balance radiative cooling at every radius but, when this is occurs, shock waves inevitably dissipate even more energy.  Our simulation explains why some clusters, such as Abell 2029, are cooling dominated, while in some other clusters, such as Perseus, various heating mechanisms including shock heating, turbulent dissipation and bubble mixing can all individually balance cooling, and together, overheat the core.

\end{abstract}


\section{Introduction}
Most relaxed galaxy clusters possess a cool-core where, in the absence of heating, rapid radiative cooling should lead to a classical cooling flow of hundreds to a thousand solar masses per year \citep{Fabian1977, Fabian1994}. However, X-ray observations by Chandra and XMM-Newton reveal a lack of cooler X-ray gas below 1-2 keV that is expected in the static cooling flow model \citep{Peterson2003}. In addition, the observed star formation rate is usually much lower than the classical rate \citep{McNamara1989, ODea2008, Hoffer2012, Donahue2015}. The absence of observational support for a classical cooling flow (the ``cooling flow problem'') suggests the presence of some heating source(s) to offset radiative cooling. It is now generally accepted that AGN feedback plays a major role in preventing a classical cooling flow \citep{McNamara2007, Fabian2012}. However, exactly how AGN deposits its energy into the intra-cluster medium (ICM) is still unclear.

Observationally, different ways of analyzing and interpreting the X-ray data have resulted in different answers to what is the major heating mechanism. Most cool-core clusters are seen to harbor X-ray cavities, which are interpreted to be AGN inflated bubbles filled with relativistic plasma and/or very hot thermal gas that are estimated to be energetically sufficient to offset radiative cooling in almost all systems \citep[e.g.,][]{Dunn2006, Hlavacek2012}. These bubbles contain abundant energy \citep{Zhuravleva2016} which can be directly mixed into the ICM, or drive shock waves, sound waves and turbulence as they rise \citep{Churazov2001}. By estimating the frequency and strength of the weak shocks in the core of the Perseus Cluster, \citet{Fabian2003} and \citet{Fabian2006} conclude that the dissipation of shock waves can offset cooling. On the other hand, \citet{Zhuravleva2014} extracts velocity spectra from the observed density fluctuations, and finds that the dissipation of turbulence can balance radiative cooling  locally at each radius within the cores of Perseus and Virgo. 

Numerical simulations and calculations have also explored a wide range of ways in which AGN heats the ICM, such as diffusion and/or streaming of cosmic rays \citep{GO2008, Enblin2011}, viscous dissipation of sound waves \citep{Ruszkowski2004}, turbulent dissipation mediated by magnetic fields and plasma instabilities \citep{Kunz2011}, hot bubbles + circulation flows \citep{Mathews2003}, and bubble mixing \citep{Hillel2016}. In recent years, momentum-driven AGN jets powered by cold mode accretion have achieved great success in reproducing many observed properties of cool-core clusters \citep{Dubois2010, Gaspari2012, PIII, Prasad2015}, including globally suppressed cooling rate, spatially extended multiphase gas that resembles the observed H$\alpha$ filaments \citep{McDonald10, PII}, weak shock waves and cavities behind shock waves, while maintaining a cool-core appearance. In addition, \citet{Li2015} showed that star formation and stellar feedback can regulate the long term AGN cycles, causing the cluster core to experience semi-periodic bursts of precipitation, star formation and AGN outbursts, which naturally explains the variety of morphologies of star forming structures observed in nearby brightest cluster galaxies (BCGs) and star formation rates \citep{Donahue2015, Tremblay2015}.

Interestingly, contrary to \citet{Zhuravleva2014}, numerical simulations of explosive AGN-like events in \citet{Reynolds2015} are inefficient in driving turbulence. Similarly, \citet{Yang2016} also concludes that turbulence is not the main source of heating in idealized simulations with momentum-driven AGN feedback. How do we reconcile simulations with observations and understand different interpretations of observations?

In this paper, using a novel method, we analyze the standard simulation in \citet{Li2015} where radiative cooling is balanced by momentum-driven AGN feedback in a time averaged sense in an isolated, idealized Perseus-like cluster. We compute the dissipative heating rate (the rate at which kinetic energy is dissipated into thermal energy), and separate it into shock dissipation and turbulent dissipation. 
We calculate the amount of heating due to different physical processes as a function of radius at different times. 
The key questions we try to address in this paper include: (1) How much energy is dissipated via shocks and turbulence? (2) How does the energy dissipation change with radius and time as the cluster core goes through cycles of AGN outbursts? (3) Do simulations agree with observations on the amount of turbulent heating in cluster cores? (4) How do we reconcile the different interpretations of observations? 

The paper is structured as follows: in Section~\ref{sec:method}, we describe the methodology including the simulation setup, the methods that we use to calculate dissipative heating rate and to separate shock dissipation and turbulent dissipation; in Section~\ref{sec:results}, we present the results of the simulation analysis, and show how cooling and heating change as a function of time and radius; In Section~\ref{sec:discussion_1} and Section~\ref{sec:discussion_2}, we discuss how our results compare with the other theoretical works and the observations. In Section~\ref{sec:discussion_3}, we discuss the uncertainties of our analysis and the limitations of momentum-driven AGN feedback model, and in Section~\ref{sec:discussion_4} we discuss the implication of our results. We summarize this work in Section~\ref{conclusions}.

\section{Methodology}
\label{sec:method}

\subsection{The Simulation}
\label{sec:method_1}

The simulation analyzed in this paper is the standard run in \citet{Li2015} which has a detailed description of the simulation setup. Here we only reiterate the key aspects.

The simulation is performed with the three dimensional adaptive mesh refinement (AMR) code Enzo \citep{Enzo}. The hydrodynamic method is the ZEUS method \citep{Zeus} which we will discuss more in Section~\ref{sec:method_2}. The resolution (smallest cell size) of the simulation is $\approx 244$ pc with a box size of 16 Mpc and a maximum refinement level of 10. 

The initial gas density and temperature are based on the observations of the Perseus cluster \citep{Churazov}. We adopt the ideal gas law with an adiabatic index of $5/3$, and assume that the ICM is initially in hydrostatic equilibrium with the gravitational potential, which includes four components: the ICM itself, the SMBH with $M_{SMBH}=3.4\times10^8 M_{\odot}$ \citep{BHmass}, the Perseus BCG NGC 1275 \citep{Mathews}, and an Navarro-Frenk-White (NFW) dark matter halo \citep{NFW} with $\rho_0 = 8.42\times 10^{-26}\, \mathrm{g}\, \mathrm{cm}^{-3}$ and the scale radius $R_s = 351.7$ kpc. 

We model the momentum-driven AGN feedback with a pair of jets (collimated bipolar outflows) launched along the z-axis from two parallel planes \citep{Omma2004}, powered by cold mode accretion, the rate of which is estimated by dividing the total amount of cold gas within the accretion zone of $r<500$ pc by a typical accretion time of 5 Myr. The jets are mass loaded such that the total outflow rate from the two jets is equal to the accretion rate $\dot{M}_{\rm SMBH}$. The jet power is $\dot{E} = \epsilon  \dot{M}_{\rm SMBH} c^2$ where we use a feedback efficiency $\epsilon=1\%$, and assume that half of the energy is thermalized at the jet launching planes. With these parameters, the jet velocity is $\sim 3\times10^4$ km $\rm s^{-1}$. We also adopt a small angle precession ($\theta=0.15$) with a 10 Myr period to avoid low density channels and increase the area that the jets directly impact \citep[e.g.,][]{Reynolds06}. 

The other important physical processes included in the simulations are radiative cooling, self gravity of the ICM, and star formation and stellar feedback. We adopt the radiative cooling curve used in \citet{Tasker2006} for half-solar metallicity gas \citep{Metallicity} with a temperature floor at 300 K. Star formation is modeled based the criteria described in \citet{CenOstriker}. A star particle is formed when the gas density exceeds a threshold ($1.67\times 10^{-24} \, \mathrm{g} \, \mathrm{cm}^{-3}$ in this simulation), and the cell is Jeans unstable with a convergent flow and a cooling time shorter than the local dynamical time. Thermal feedback from star particles is also included. Though as discussed in \citet{Li2015}, stellar feedback has little impact on the results because its power is orders of magnitudes lower than that of AGN feedback in a cluster.

\subsection{Dissipative Heating}
\label{sec:method_2}

One main goal of this work is to study how mechanical energy of the jets is dissipated into heat. We chose the ZEUS \citep{Zeus} hydrodynamic method for the simulation, which uses a von Neumann-Richtmyer artificial viscosity, enabling us to calculate the rate at which kinetic energy is dissipated in the simulations. 

The details of the ZEUS method can be found in \citet{Zeus, Zeus2, Enzo}. We only present the key formula that is relevant to the calculation. Zeus method employs an artificial viscous pressure to smooth shock discontinuities:

\begin{equation}
\label{eq:Zeus}
q=\begin{cases}l^2\rho(\triangledown\cdot v)^2  &if \, \triangledown\cdot v < 0, \\0 &otherwise,\end{cases}
\end{equation}

where $l$ is a constant that determines the strength of the viscosity, and is set to be $\sqrt{2}\Delta x$ in the simulation with $\Delta x$ being the cell width. At every time step of the simulation, where the condition $\triangledown\cdot v < 0$ is met, a viscous heating term $-q(\triangledown \cdot v)$ is added to the energy equation, which is the kinetic energy that is converted into thermal energy locally. We show how this method works in a test simulation in the Appendix.

\subsection{Turbulence vs. Shocks}
\label{sec:method_3}

The method described in Section~\ref{sec:method_2} allows us to compute the total dissipative heating rate within each cell at each time step. We then separate shock heating and turbulent heating by identifying the cells that harbor shock waves. 

Various methods have been used in the literature \citep[e.g.][]{Ryu2003, Skillman2011} to identify shock waves based on the jump conditions. We adopt the shock finding method used in the Enzo refinement criteria. A cell is flagged to contain shocks if all of the following conditions are met: (1) $P_2/P_1>1+\epsilon_P$, where $P_2/P_1$ is the maximum pressure jump across any interface of that cell.
$\epsilon_P$ is a small positive value to ensure that pressure increases after the shock. We set $\epsilon_P=1.002$ which corresponds to a Mach number of $\sim 1.001$. Our results are not sensitive to the exact value of $\epsilon_P$ as we discuss in Section~\ref{sec:discussion_3}. (2) $\triangledown \cdot v < 0$, i.e., the flow is converging. (3) The thermal energy to total energy ratio within the cell $e_{thermal}/e_{total}>0.1$. We have also experimented using a temperature criterion that requires the gradients of gas temperature and entropy to have the same sign \citep{Ryu2003, Skillman2011}, and find little difference in the results.

After we flag all the cells that have shocks, we assume that the flagged cells are dissipating kinetic energy via shock waves, whereas the rest of the cells that are dissipating kinetic energy are doing so via turbulence. Because turbulence can also exist in cells that have shocks, the amount of turbulence is likely underestimated, but since the volume filling fraction of shock waves is rather low 
compared with turbulence, the amount of turbulence that is unaccounted for should be very small.

\begin{figure*}
\begin{center}
\includegraphics[scale=.27]{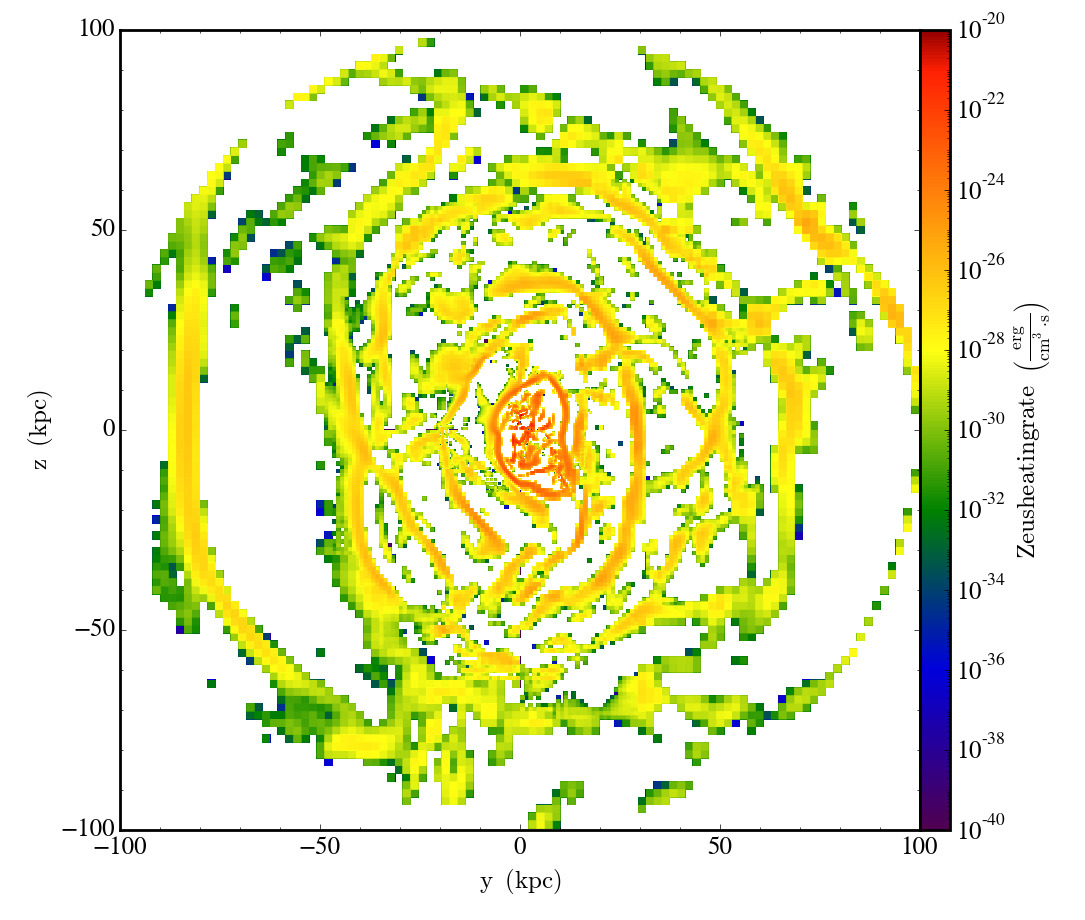}
\includegraphics[scale=.27]{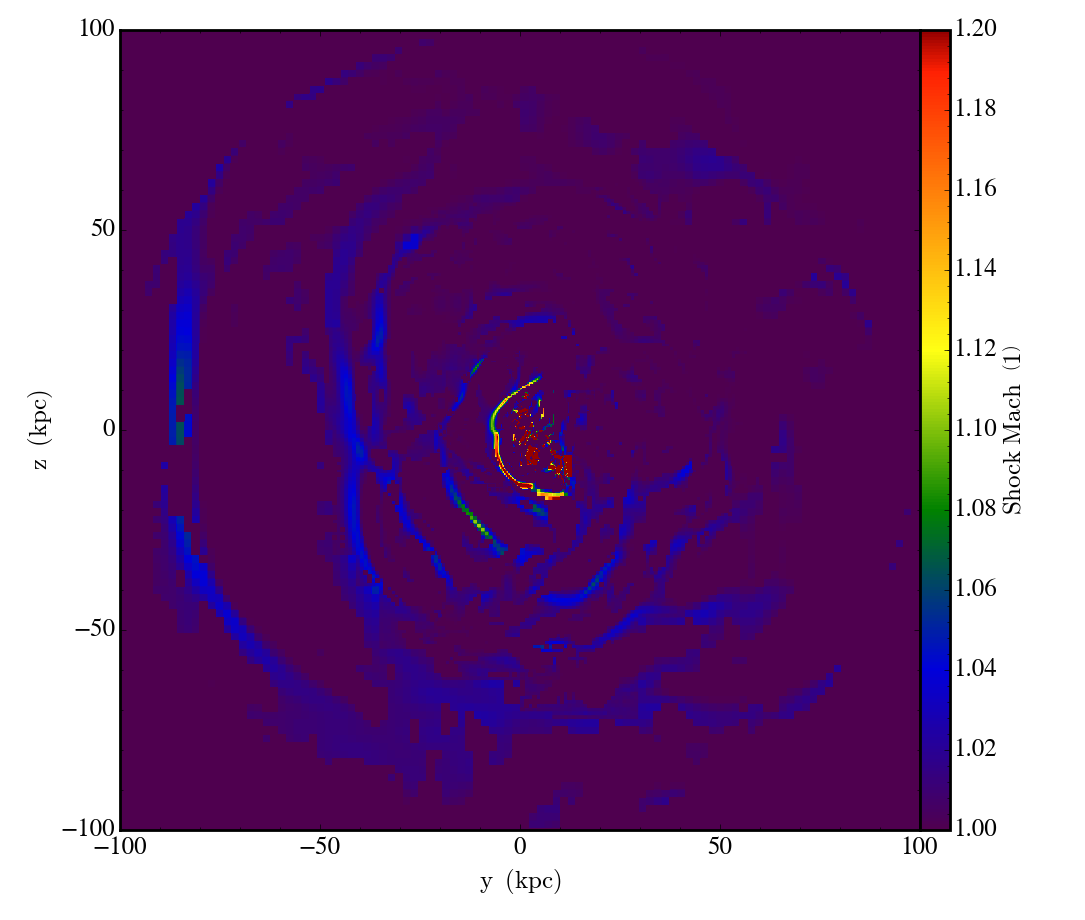}

\caption{Dissipative heating rate (left) and the Mach number of the shock waves (right) in a slice through the center of the cluster along the y-z plane at $t=0.88$ Gyr. 
\label{fig:slice}}
\end{center}
\end{figure*}

Figure~\ref{fig:slice} shows the dissipative heating rate and the Mach number of the shock waves in a slice of gas through the center of the cluster along the y-z plane at $t=0.88$ Gyr. Note that only gas at $r < 100$ kpc (roughly the cooling radius of the cluster) is shown. We also excluded star forming regions to avoid contamination by selecting only gas with temperatures higher than $10^6 K$ and densities lower than $10^{-23} g/cm^3$.

\section{Results}
\label{sec:results}
\subsection{Heating and cooling as a function of time}

The cluster experiences three cooling-AGN outburst cycles within $\sim 6.5$ Gyr. The details of the interplay between ICM cooling, AGN feedback, and star formation are described in \citet{Li2015} with a focus on the role of star formation and stellar feedback. In this section, we examine the battle between radiative cooling and AGN heating.

\begin{figure}
\begin{center}
\includegraphics[scale=.5]{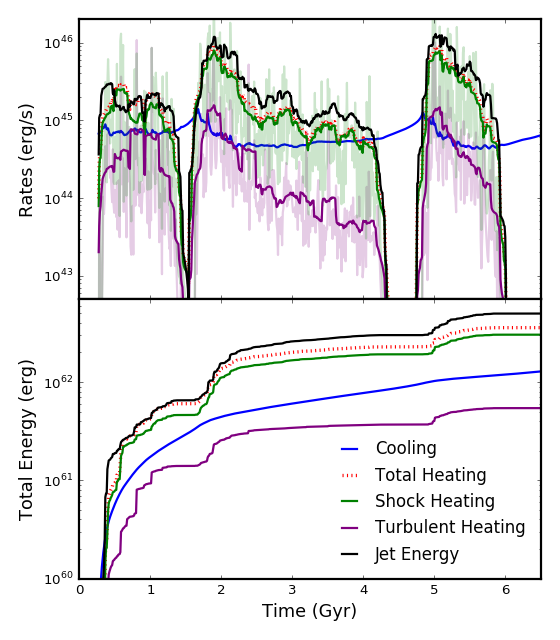}
\caption{Top: The radiative cooling rate from all the gas within $r<100$ kpc (solid blue line) as a function of time, along with the jet power (solid black at the top) smoothed with a moving window of 200 Myr, the smoothed total dissipative heating rate (dotted red line), the smoothed shock dissipation rate (solid green line) and the turbulent dissipation rate (solid purple line at the bottom). The original un-smoothed data are shown in semi-transparent colors for shock heating and turbulent heating. Bottom: the total cumulative energy loss within $r<100$ kpc due to radiative cooling as a function of time, compared with the total energy output from the AGN, and the total energy dissipated within $r<100$ kpc, which has two components--shock heating (green) and turbulent heating (purple). 
\label{fig:combine2}}
\end{center}
\end{figure}

The top panel of Figure~\ref{fig:combine2} shows the evolution of the AGN jet power, along with the ICM cooling rate, the total dissipative heating rate, shock heating rate and turbulent heating rate within $r < 100$ kpc. The jet power and all the heating rates show very large variations on short timescales (the semitransparent lines show the original data for the shock heating rate and the turbulent heating rate). For clarity, we only plot the time averaged jet power and the total dissipative heating rate.

At the beginning of each cycle, right before AGN feedback is triggered, there is no heating, and therefore the ICM cooling rate increases as the cluster core becomes cooler, trying to form a classical cooling flow. As soon as global cooling happens at the very center of the cluster, the cold-mode accretion turns on AGN jets, which quickly trigger more ICM to precipitate as the jets dredge out low entropy gas to larger radii \citep{PII, Mark2016}. The cold gas that condenses out of this non-linear thermal instability partially forms stars and partially falls onto the SMBH, boosting its feedback power. The jet power drastically increases due to this short period of stimulated precipitation (``positive AGN feedback''), and goes hand-in-hand with the shock dissipative rate, followed by turbulent dissipation with a delay of about 100 Myr. Within $\sim200-300$ Myr, the AGN power reaches its peak. The ICM cooling rate peaks before that and begins to decrease as AGN heating (``negative AGN feedback'') starts to take effect. 

After the AGN outburst (accompanied by a burst of star formation), heating continues to dominate cooling. The reduced cooling rate causes a gradual decline in the AGN power. Towards the end of the cycle, as star formation eventually consumes all the cold gas, taking away the fuel for the SMBH, AGN feedback is shut off. The jet power and all the heating rates go to zero, allowing the ICM cooling rate to increase again and the system to enter the next cycle.

The shock heating rate is almost an order of magnitude higher than the turbulent heating rate, and is therefore very close to the total dissipative heating rate. The turbulent heating rate is almost always below the cooling rate except during AGN outbursts, which we will discuss more in Section~\ref{sec:discussion_1} and ~\ref{sec:discussion_2}.

The bottom panel of Figure~\ref{fig:combine2} shows the cumulative energy output from the AGN, and different heating processes, along with the cumulative energy loss due to radiative cooling within $r < 100$ kpc. At the end of the simulation, the total energy output from the AGN is $\sim 5\times10^{62}$ erg, about $60\%$ of which is dissipated via shock waves, and $10\%$ via turbulence. The rest $30\%$ is dissipated outside of the $r=100$ kpc core region. Not all of the thermal energy is confined within $r<100$ kpc either. The total energy loss due to radiative cooling is $1.3\times 10^{62}$ erg, only $1/3$ of the thermal energy that is generated by the AGN jets within $r=100$ kpc. Given that the core roughly returns to its original state at the end of the simulation, $2/3$ of the thermal energy is transported to $r>100$ kpc. 

\begin{figure}
\begin{center}
\includegraphics[scale=.38]{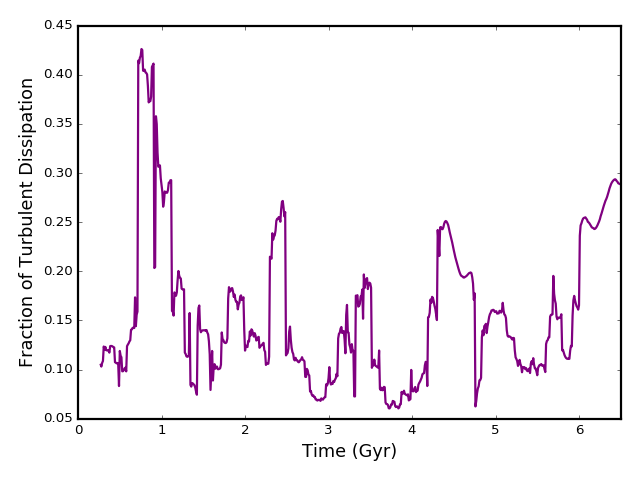}
\caption{The fraction of turbulent dissipation within $r<100$ kpc as a function of time. 
\label{fig:Turb_fraction}}
\end{center}
\end{figure}

Although the absolute amount of shock and turbulent dissipation correlates with AGN cycles (Figure~\ref{fig:combine2}); the relative amount of heating due to weak shocks vs. turbulence does not appear to correlate with the cycles in an obvious way as shown in Figure~\ref{fig:Turb_fraction}. The fraction of turbulent dissipation within $r<100$ kpc varies from 5\% up to 40\%. Although some peaks appear to follow major AGN outbursts (e.g. the first peak at around 1 Gyr), most of them do not. The lack of a clear correlation is likely because the relative importance of turbulent dissipation can be affected by the interaction between AGN jets and cold clouds, which is stochastic.

\subsection{The radial dependence of heating and cooling}

\begin{figure}
\begin{center}
\includegraphics[scale=.4]{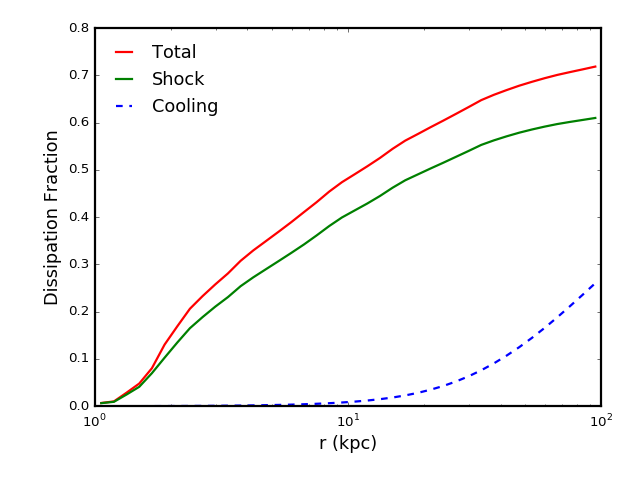}
\caption{The fraction of the jet energy that is dissipated as a function of radius (solid red line) averaged over the whole duration of the simulation. The solid green line shows the shock dissipation fraction, and the dashed blue line shows the required dissipation rate if dissipative heating perfectly balances cooling at every radius.
\label{fig:All_average_r}}
\end{center}
\end{figure}

Figure~\ref{fig:All_average_r} further shows the imperfection of AGN heating: less than $80\%$ of the total jet energy is dissipated within the core, and the amount of energy that is dissipated far exceeds what is required to perfectly balance the radiative cooling loss. The coupling of the AGN jets to the ICM is not perfect. Thus in order to suppress cooling flows, the SMBH needs to inject a few times more energy than what is needed to offset cooling.

\begin{figure*}
\begin{center}
\includegraphics[scale=.7]{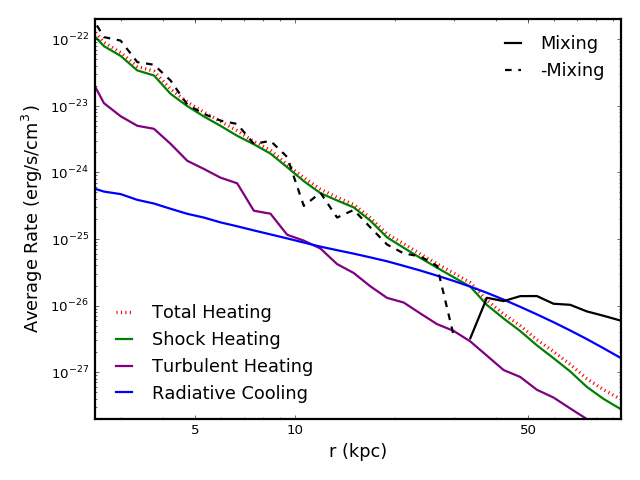}\\
\includegraphics[scale=.3,trim=0cm 0cm 0.62cm 0cm, clip=true]{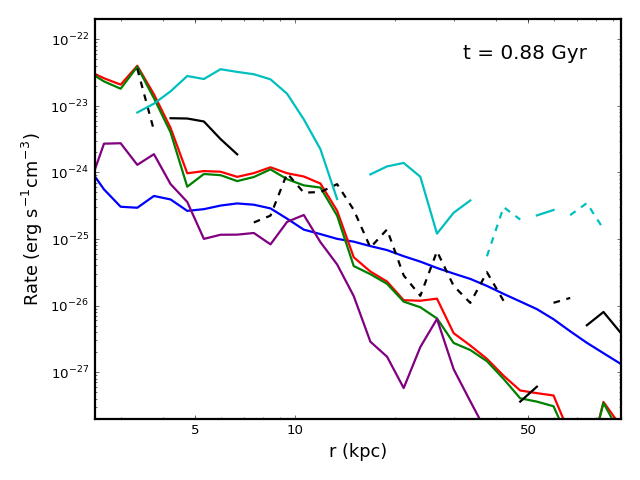}
\includegraphics[scale=.3,trim=3cm 0cm 0.62cm 0cm, clip=true]{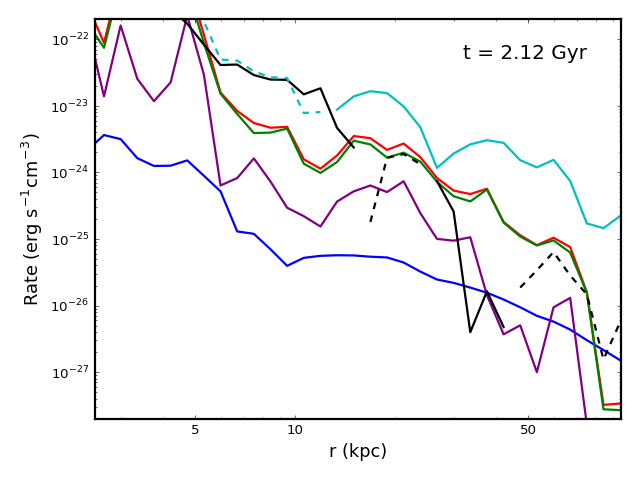}
\includegraphics[scale=.3,trim=3cm 0cm 0.62cm 0cm, clip=true]{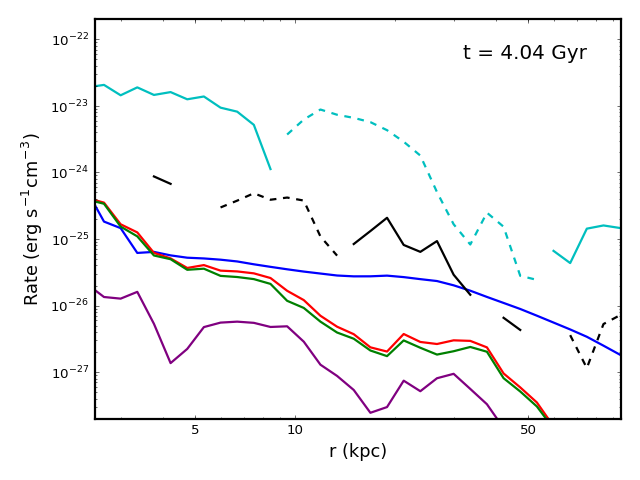}
\caption{Top: The average radiative cooling rate (solid blue line), total dissipative heating rate (dotted red line), shock heating (solid green line) and turbulent heating (solid purple line) rate as a function of radius weighted by cell volume. Bottom: Cooling and heating rates at three different times. The black lines show the rate at which thermal energy is transferred through radial mixing, and the cyan lines are the change of local thermal energy due to adiabatic expansion and compression. The dashed parts are showing the negation of the rate (see Section~\ref{sec:results} for how these rates are calculated). \label{fig:mean_r}}
\end{center}
\end{figure*}

Even within the core, radiative cooling is not perfectly balanced by the local dissipative heating at every radius. Figure~\ref{fig:mean_r} shows cooling and heating rates as a function of radius averaged over the whole duration of the simulation (top panel) and at three representative times (bottom three panels). The dissipative heating rates decrease quickly with radius, roughly with a power law slope of $r^{-3.1}$, compared with the cooling rate which has a shallower slope of $r^{-1.2}$ within $r=30$ kpc. There is more dissipative heating than radiative cooling within $r \sim 30$ kpc, and less at larger radii, indicating that radial heat transportation is important. 

To compute the rate at which heat is transported radially, we divide the cluster core into 40 spherical shells. Within each shell, the change in the total internal energy follows

\begin{equation}
\frac{\partial E_{th}}{\partial t} +v\cdot \nabla E_{th}+P\nabla\cdot v = \mathcal{H}-\mathcal{C}\, ,
\end{equation}
where $\mathcal{C}$ is the radiative cooling rate, and $\mathcal{H}$ is the dissipative heating rate which includes shock heating and turbulent heating that are discussed previously. The second term on the left hand side of the equation, $v\cdot \nabla E_{th}$, is the advection of heat (radial mixing). The last term $P\nabla\cdot v$ is adiabatic processes (expansion and compression). Because of the cyclical behavior of the cluster, the cumulative effect of adiabatic processes is negligible. Therefore, averaged over time, the radial mixing rate within each shell can be estimated as:
\begin{equation}
\langle \dot {E}_{mixing} \rangle = \langle v\cdot \nabla E_{th}\rangle \approx  \langle \mathcal{H}\rangle - \langle\mathcal{C}\rangle - \langle \frac{\Delta E_{th}}{\Delta t}\rangle \, ,
\end{equation}
where $\Delta t$ is the time interval between two simulation outputs and $\Delta E_{th}$ is the change in thermal energy in that shell within $\Delta t$. The average mixing rate is plotted as the black line in the top panel of Figure~\ref{fig:mean_r}. The dotted part denotes the negation of the rate, which almost overlaps with the total dissipative heating rate in the innermost core region ($r < 20$ kpc). This means that most of the heat dissipated within $r < 20$ kpc is transported outwards, and only a small fraction is used to offset radiative cooling. 

Even though the average effect of adiabatic processes is small, at individual time steps, the change of local thermal energy due to adiabatic expansion/compression can be quite significant (bottom panels of Figure~\ref{fig:mean_r}), often even more than radial mixing. The adiabatic processes and radial mixing of heat can be intuitively understood by examining the evolution of gas temperature in the central region of the cluster:  \url{https://vimeo.com/117832510}. Figure~\ref{fig:Tproj} shows one snapshot at $t=1.73$ Gyr, when the cluster is entering the second cycle of AGN outburst. Following AGN-driven shock waves, pockets of hot gas (low density bubbles) are being transported outwards due to both buoyancy and the positive radial post-shock velocity. These bubbles push the surrounding gas (adiabatic processes) and also mix with it.

Figure~\ref{fig:mean_r} shows that even though most of the energy is dissipated via shock waves, the importance of turbulent dissipation increase with radius. This can be seen more clearly in Figure~\ref{fig:Heatingfraction} which shows the fractional contribution from different processes to the total dissipative heating as a function of radius. Turbulent heating increases from $\sim 10\%$ to almost $40\%$ at $r\sim 100$ kpc. When we further divide shock waves in two groups based on their Mach numbers, we find that even though most of the shock waves are very weak with Mach numbers smaller than 1.1, a significant fraction of the energy is dissipated via stronger shocks with Mach numbers larger than 1.1 (see Figure~\ref{fig:slice}). These stronger shock waves dissipate more quickly as one would expect. 

The left panel of Figure~\ref{fig:Mach} further shows that most shock waves have Mach numbers below 1.2-1.3. At radii larger than a few tens of kpc, shocks with Mach numbers higher than 1.5 are extremely rare. Though as the right panel of Figure~\ref{fig:Mach} shows, even though weak shocks are abundant, the contribution to dissipative heating from strong shocks at small radii is significant.

\section{Discussion}
\label{discussion}
In this section, we first compare our results with other theoretical and numerical works in Section~\ref{sec:discussion_1}. We then compare the turbulent heating rate in the simulations with the observations in Section~\ref{sec:discussion_2}, where we also compare our simulations with the Hitomi observations of Perseus. In Section~\ref{sec:discussion_3}, we discuss the limitations of our model and in Section~\ref{sec:discussion_4}, the implications of our analysis.

\subsection{Comparison with Other Works}
\label{sec:discussion_1}
Heating cool-core clusters with weak shock waves has been extensively discussed in the literature \citep[e.g.,][]{Ruszkowski2004, VoitDonahue2005, Fabian2005, Mathews, Sternberg2009}. All of these works generally agree that shock heating can be energetically sufficient to offset cooling. By analyzing the ripples observed in the core of the Perseus Cluster, \citet{Fabian2003} and \citet{Fabian2006} conclude that heating from weak shocks and sounds waves can balance cooling. Similarly, \citet{PIII} finds that shock heating can provide enough energy by measuring the typical strength and separation of shock waves seen in the simulations.

However, shock heating has a steeper radial profile than cooling as is shown in Figure~\ref{fig:mean_r}. \citet{Fabian2005} provides an analytical expression for the dissipation rate of weak shocks as a function of radius and thermal properties of the ICM (which are also functions of radius) (see Equation 20 of the paper). For the density and temperature profiles of Perseus, they estimate that at about $10-20$ kpc, the heating rate $\epsilon_{shock} \propto r^{-3}$, in excellent agreement with the slope we find in our simulation ($\epsilon_{shock} \propto r^{-3.1}$, see Section~\ref{sec:results} and Figure~\ref{fig:mean_r}). Using 1D calculations, \citet{Mathews} shows that shock waves can deposit too much energy near the center of the cluster, causing the temperature of the ICM to rise far above the observations. They argue that this can be avoided when circulation flows are considered \citep{Mathews2003}. 

A similar idea is explored in \citet{VoitDonahue2005} which essentially describes what is seen in our simulations: the inner parts of the cluster core $r < 30$ kpc is shock heated by intermittent AGN outbursts; the outer parts of the cluster core, where shock waves have decayed, are heated by buoyant bubbles; between heating episodes, the core relaxes towards an asymptotic pure cooling profile. 

\begin{figure}
\begin{center}
\includegraphics[scale=.48]{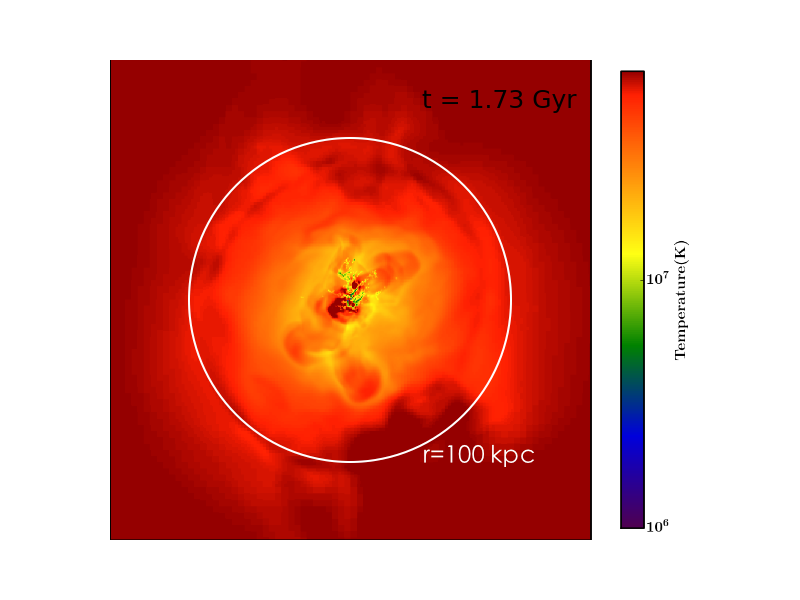}
\caption{The projected gas temperature in the central region of the cluster at $t=1.73$ Gyr. Hot bubbles (pockets of shock heated gas) are moving outwards behind shock waves, transporting the thermal energy dissipated at small radii to larger distances, some of which is transported beyond $r=100$ kcp along with a fraction of the jet energy that is still in the kinetic form.
\label{fig:Tproj}}
\end{center}
\end{figure}

The importance of bubble heating is also discussed in \citet{Zhuravleva2016, Hillel2016, Yang2016}. Our simulation setup is very similar to \citet{Yang2016} with momentum-driven AGN jets in an idealized cool-core cluster. The main difference between the two simulations is the treatment of cold gas that condenses out of the ICM. In \citet{Yang2016}, cold gas is replaced by passive particles and is assumed to be all accreted onto the SMBH. The system quickly settles to a quasi-steady state, where AGN heats and pushes the gas within the jet cones, and the ambient gas returns in a reduced cooling flow, forming a circulation. In our simulation, cold gas is preserved and is allowed to form stars. As a result, in addition to short time variabilities, AGN also undergoes Gyr timescale cycles regulated by itself and star formation. Another effect of the cold gas is that it can block and redirect the AGN jet material, changing the outflow direction stochastically, and thus causing the heating to happen in a more isotropic fashion (see \citet{PIII} and Figure~\ref{fig:Tproj}). Although our simulation does not always have an ordered circulation flow, we find that mixing and adiabatic processes are very important in redistributing the energy from the AGN, in agreement with \citet{Yang2016}.

One important conclusion in \citet{Yang2016} is that turbulent heating is unimportant in the simulation because the total energy in turbulence is small compared with thermal energy. This is in agreement with \citet{Reynolds2015} which, by decomposing the velocity field, finds that AGN is inefficient in driving turbulence and less than $1\%$ of the injected energy ends up in the turbulence. In our simulations, about $10\%$ of the energy is dissipated via turbulence. AGN jets seem more efficient in driving turbulence in our simulations, likely due to the different treatment of the cold gas discussed earlier, which results in a randomization of the jet direction instead of a more steady, ordered ``circulation'' in \citet{Yang2016}. Nonetheless, in agreement with other simulations, our results also suggests that turbulent dissipation is far from the dominant heating mechanism. Thus numerical simulations appear to contradict \citet{Zhuravleva2014} which finds that turbulent heating can compensate for radiative cooling at every radius within the cores of Perseus and Virgo. 

\begin{figure}
\begin{center}
\includegraphics[scale=.38]{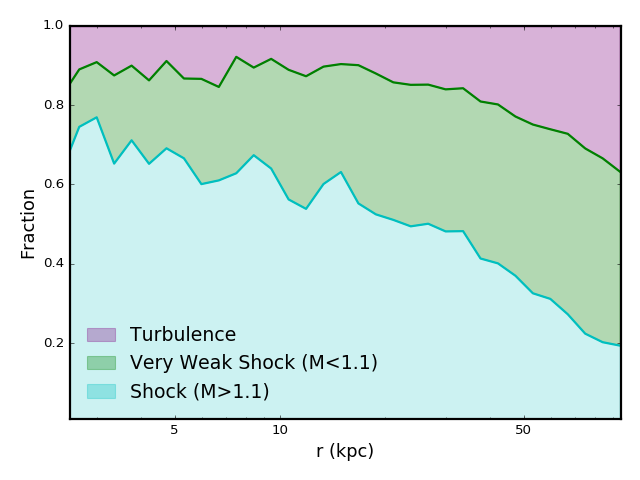}
\caption{The fraction of energy that is dissipated via turbulence (purple) vs. shock waves (green). Shock dissipation is further divided based on the Mach number of the shock waves with lighter green denoting stronger shocks. As one would expect, stronger shock waves dissipate faster. The significance of turbulent dissipation increases with radius, but remains subdominant within the core.  
\label{fig:Heatingfraction}}
\end{center}
\end{figure}

\begin{figure*}
\begin{center}
\includegraphics[scale=.4]{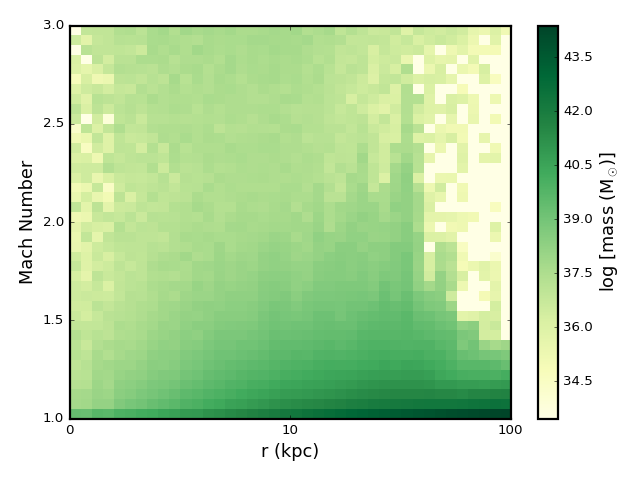}
\includegraphics[scale=.4]{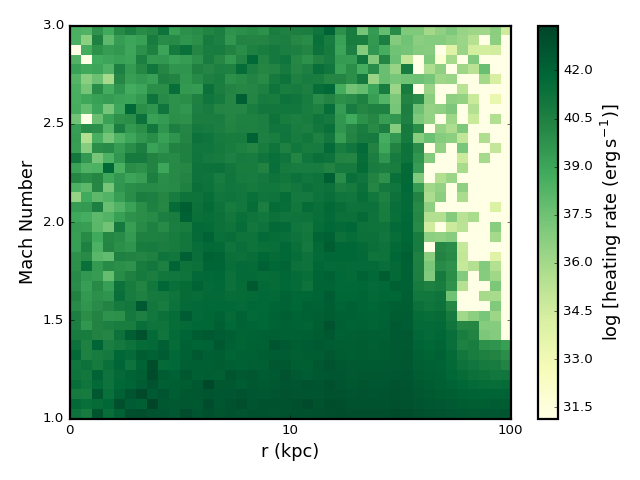}
\caption{The distribution of shock Mach number as a function of radius averaged over the entire simulation, weighted by mass (left) and dissipative heating  rate (right). Very weak shock waves with Mach numbers below 1.5 are most abundant, but the stronger shocks contribute a lot to the total dissipative heating. \label{fig:Mach}}
\end{center}
\end{figure*}

\subsection{Comparison with Observations}
\label{sec:discussion_2}
In this section, we discuss how the simulation results compare with the observations. We focus on the amount of turbulent dissipation vs. radiative cooling in Section~\ref{sec:discussion_2a}. In Section~\ref{sec:discussion_2b}, we discuss the velocities of the ICM in comparison with the Hitomi observations of Perseus \citep{Hitomi2016}, and in particular, why our weak shocks do not contribute much to the line broadening.

\subsubsection{Turbulent Heating Rate}\label{sec:discussion_2a}
\begin{figure}
\begin{center}
\includegraphics[scale=.4]{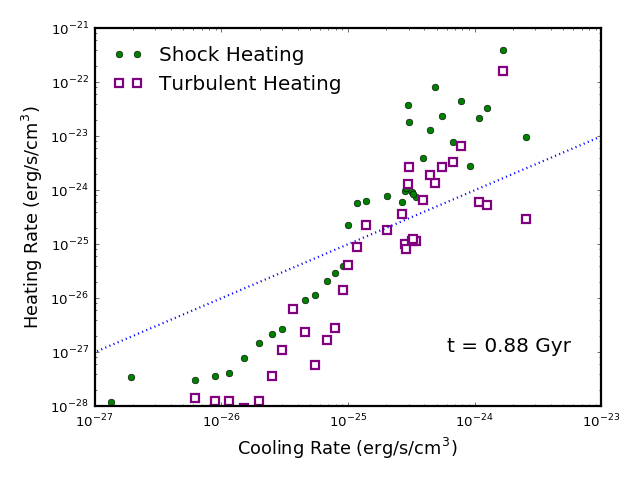} 
\caption{The radiative cooling rate inside each radial bin of the cluster core compared with the local turbulent heating rate (purple squares) estimated directly from the simulation data as described in Section~\ref{sec:results}. The dotted line denotes the 1:1 ratio. Turbulent heating and radiative cooling seem well balanced, in agreement with the observations of Perseus and Virgo \citep{Zhuravleva2014}, but at the same time, there is even more dissipative heating from shock waves (green circles). 
\label{fig:Yuan}}
\end{center}
\end{figure}

To understand the conflict between the observations and numerical simulations on turbulent heating, we compare the amount of heating and cooling in the simulation at every radius, similar to \citet{Zhuravleva2014}. We divide the cluster core into 40 shells and plot the radiative cooling rate against turbulent heating and shock heating rates within each shell in Figure~\ref{fig:Yuan} for t=0.88 Gyr. The blue solid line shows the 1:1 ratio. Above $10^{-25} erg/s/cm^3$, the squares closely follow the blue line, suggesting that turbulent heating can roughly balance cooling within every shell except in the outer core region. However, Figure~\ref{fig:Yuan} also shows that although there is enough heating from turbulence dissipation, there is yet more from shock wave dissipation. Meanwhile, as the bottom left panel of Figure~\ref{fig:mean_r} shows, even more heat is transported radially. In fact, at $t=0.88$ Gyr, the total dissipative heating rate exceeds the radiative cooling rate (see Figure~\ref{fig:combine2}): the cluster core is being over-heated.

Indeed, as Figure~\ref{fig:combine2} and ~\ref{fig:mean_r} show, the cluster is not in a steady state where AGN heating perfectly balances radiative cooling. Rather, the cluster core is over-heated following each AGN outburst, and is under-heated at the very beginning and the end of each cycle. According to our model, when the cluster is over heated, it usually has extended multi-phase gas, enhanced star formation and AGN activities (although the AGN radio power may not be particularly high observationally due to the large temporal variation of cold mode accretion onto the SMBH). Whereas, an under-heated cluster core is typically seen with low AGN power and not much cold gas or star formation activities. 

The Perseus cluster has extended multi-phase gas \citep[e.g.,][]{Conselice2001}, a strong nuclear X-ray source with $L_{X,nuc}=25^{+54}_{-17}$ keV \citep{Merloni2007} and a rather high far-infrared SFR of $24\pm1 M_{\odot} yr^{-1}$ \citep{Mittal2012}. Therefore, it is likely to be in the evolutionary stage when heating overwhelms cooling, and thus the seemingly conflicting claims based on the observations of Perseus can be reconciled: there is indeed enough energy from the dissipation of turbulence \citep{Zhuravleva2014} or shock waves alone \citep{Fabian2003, Fabian2006} to offset cooling, and there is also abundant (possibly even more) energy in the hot bubbles \citep{Zhuravleva2016}. All heating processes exist at the same time and over-heat the cluster core. Some of the thermal energy escapes the inner core region, while some increases the core entropy, which slows down precipitation and causes the average AGN power to decline. Gradually, heating loses to cooling, causing the core entropy to decrease. The core relaxes towards a classical cooling flow, and eventually, global catastrophic cooling in the center of the cluster occurs, and triggers another AGN outburst.

\subsubsection{Velocities of the ICM}\label{sec:discussion_2b}
The Hitomi X-ray Observatory recently took high-resolution spectra of the Perseus core, and the line widths show a rather low line-of-sight velocity dispersion of $164 \pm 10 km/s$ in a region $30-60$ kpc from the central nucleus, suggesting a ``quiescent ICM'' \citep{Hitomi2016}. This is in concordance with our simulations where turbulent dissipation is not significant. We also argue that even though shock waves are the main channel to dissipating kinetic energy in our simulations, they may not contribute much to the line broadening. The reasons are: (i) Strong shocks that provide most of the heating can be concealed because there are only a few of them (Figure~\ref{fig:Mach}). They are not volume filling, and the line emission from them is ``buried'' by the emission from the slower gas located in between them. (ii) Weak shocks do not broaden the lines significantly because the associated line shifts are much smaller than the sound speed $c_s$. For weak shocks, the post-shock velocity in the shock frame is $(1+\frac{3}{2}\epsilon) v_{shock}$, where $\epsilon=M-1$, and $v_{shock}=(1+\epsilon)c_s$. Thus to first odder accuracy, the post-shock velocity that causes the line shift in the observer's frame is 
\begin{equation}
v_{shift}=\frac{3}{2}\epsilon c_s\,. 
\end{equation}
Most shock waves in the simulation have a Mach number of $M=1.1$ or smaller. For $M=1.1$ and $c_s=800$ km/s, $v_{shift}=120 \,\rm km/s$, which is consistent with the Hitomi data.

\begin{figure*}
\begin{center}
\includegraphics[scale=.4]{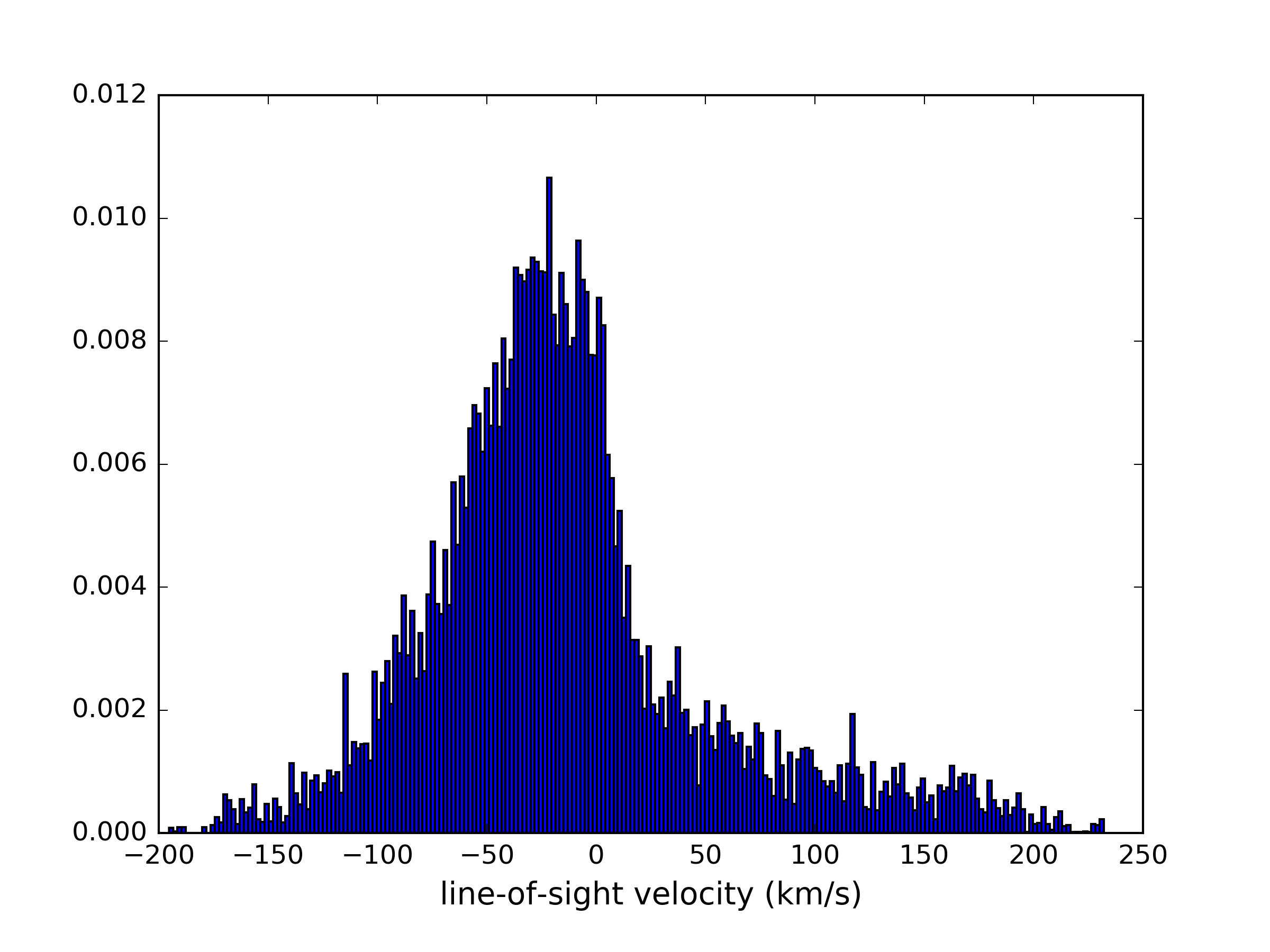}
\includegraphics[scale=.4]{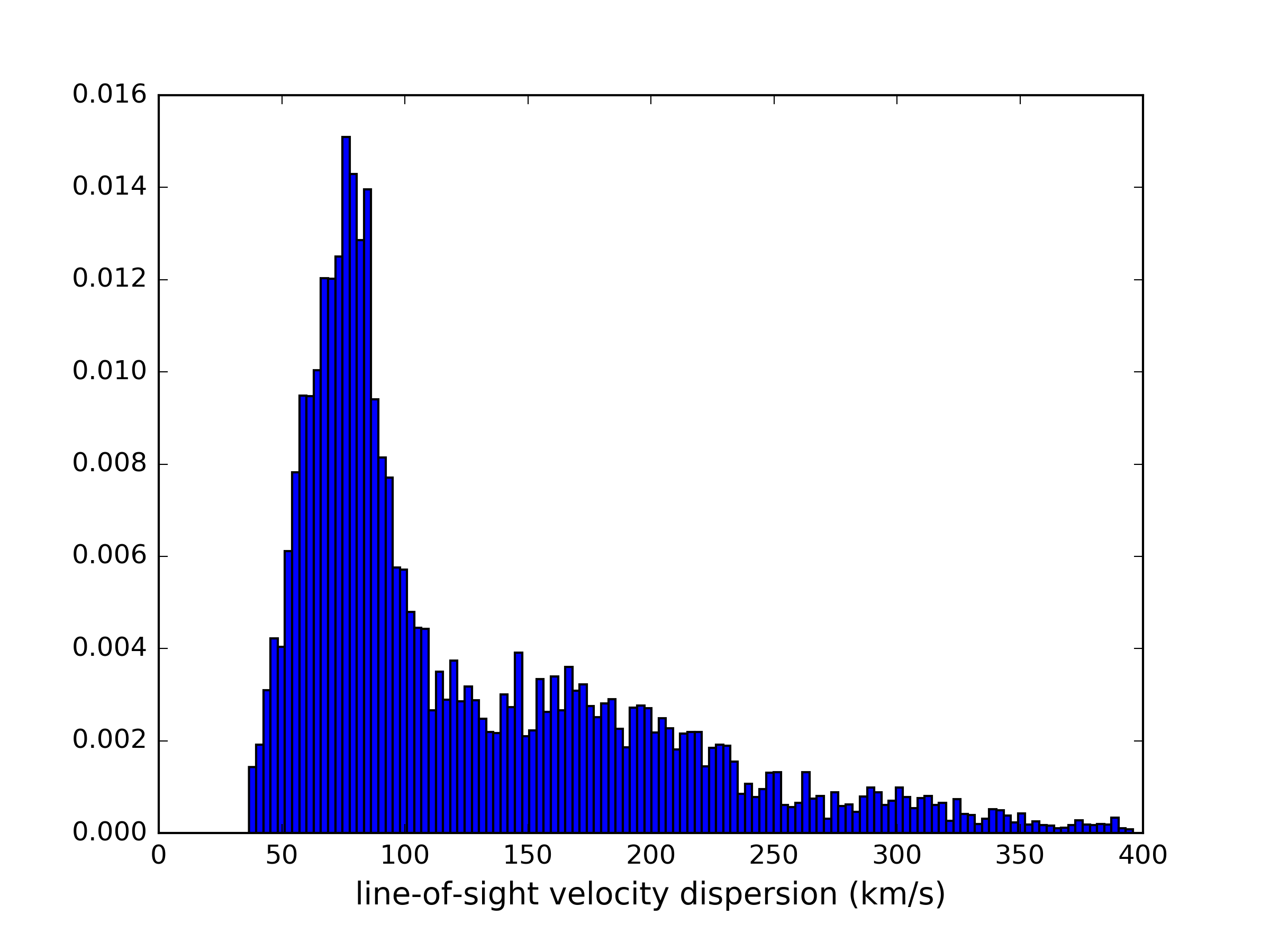} 
\caption{Left: the distribution of the average line-of-sight velocities viewed along the x-axis at t=0.88 Gyr within radius of 30-60 kpc. Right: the distribution of the velocity dispersions along individual lines of sight. 
\label{fig:Hitomi}}
\end{center}
\end{figure*}

The left panel of Figure~\ref{fig:Hitomi} shows the distribution of the average line-of-sight velocities measured along the x-axis in the simulation at t=0.88 Gyr. The region is selected to be a cylinder of radius 60 kpc with the central 30 kpc cut out to compare with the Hitomi observations, which is a region $30-60$ kpc from the central nucleus. Even though shock heating and turbulent dissipation are over-heating the cluster core at this moment, most of the gas has velocities lower than 200 km/s. The right panel shows the distribution of the velocity dispersion along individual lines of sight. Most area has line-of-sight velocity dispersions around $50-100$ km/s. We plan to compare with the Hitomi observations in more detail in future work (preliminary results confirm consistency). 

\subsection{Limitations}
\label{sec:discussion_3}

In this section, we first discuss the limitations of our model, including the uncertainties of the analysis, and the limitation due to important physical processes that are not considered in the simulations. Lastly we discuss the implications of our results. 

One of the main uncertainties in our analysis comes from the conservation of energy when using Zeus method. Because Zeus follows only the thermal energy, not the total energy, the total energy may not be conserved. As we show in the Appendix, energy loss can be as high as ($\sim 30 \%$) in a low resolution test of decaying turbulence with fixed grids and high Mach numbers. Because the simulation analyzed in this paper uses AMR and the Mach numbers are typically rather low ($<1.5$, see Figure~\ref{fig:Mach}), the energy loss is likely less severe. The fact that the sum of shock and turbulence losses within $r<100$ kpc are close to the energy input is telling us that energy non-conservation is unlikely to be a big effect here (Figure~\ref{fig:combine2} and Figure~\ref{fig:All_average_r}). The effect of energy non-conservation in the simulation is equivalent to a slightly enhanced cooling rate, and results in a slightly higher burden on the AGN. However, we note that because of the energy loss, the feedback energy that escapes beyond $r=100$ kpc in the simulation may be slightly lower than what is inferred in our analysis. Future work should address this, for example, by comparing simulations with Zeus and other energy-conserving methods.

Another uncertainty comes from how we distinguish shock waves and turbulent motion. We use $\epsilon_P=1.002$ in shock identification, which corresponds to an extremely low Mach number of $\sim 1.001$. To test how sensitive our results are to the choice of this parameter, we analyze the simulation data using $\epsilon_P=1.02$, corresponding to a Mach number of $\sim 1.01$. We find that the total energy dissipated by shockwaves within $r<100$ kpc only changes by $\sim 1\%$. In other words, extremely weak shock waves with Mach numbers of 1.001-1.01 only dissipate 1\% of the total energy. This is in line with our results in Section~\ref{sec:results} showing that weaker shocks are less effective at dissipating energy (Figure~\ref{fig:Heatingfraction} and Figure~\ref{fig:Mach}) even though they are abundant in the simulation. 

The results in this paper come from AGN feedback modeled using pure hydrodynamic simulations. Like the other pure hydro simulations with momentum-driven AGN feedback, the bubbles/cavities here are filled with non-relativistic gas that is shock heated by the fast moving AGN jets. In reality, the shocks are likely driven by the expansion of the bubbles that contain relativistic plasma. The mass loaded jets in simulations do slow down very quickly. At a few kpc to a few tens of kpc, the typical Mach numbers of the shocks are $\sim 1.1-1.2$ (Figure~\ref{fig:slice}, see also \citet{Gaspari2011,PIII}), generally consistent with the weak shocks observed in nearby clusters \citep{Fabian2006, Blanton2011}. However, the hot bubbles in the hydro simulations usually lack the smooth spherical shape and break easily whereas the observed bubbles appear to be longer lived. The solution may be adding viscosity \citep{Reynolds2005} or magnetic fields \citep{Lyutikov2006, Ruszkowski2008} to the simulations. In addition, cluster weather due to mergers and sloshing \citep[e.g.][]{ZuHone2010} may source turbulent motion or even facilitate precipitation, and these effects are not included in the idealized simulations here.

\subsection{Implications}
\label{sec:discussion_4}

Despite the lack of physical processes discussed previously, our model seems to produce results that are in general agreement with the observations in terms of the amount of shock heating and turbulent heating: turbulence may sometimes generate a considerable amount of heating compared with cooling \citep{Zhuravleva2016}, but turbulent heating is almost always subdominant compared with shock heating. Even though the exact duration and strength of each AGN cycle may be sensitive to the simulation parameters such as the star formation efficiency and the modeling of stellar feedback, the cyclical behavior itself is robust. This explains why in some cool-core clusters, AGN feedback appears insufficient, such as Abell 2029 \citep{Rachel2013}, while some other clusters are observed to be heating dominated \citep[e.g.][]{McNamara2005}. Additionally, a considerable fraction of the energy (both thermal and kinetic) can escape the innermost core region during the peak of the heating dominant phase in our model, which corresponds to clusters like Hydra \citep{Nulsen2005, Wise2007}. Note that even though the jet power and the heating rates show large swings with time (Figure~\ref{fig:combine2}), the radiative cooling rate of the ICM exhibits much more subtle variations. 
We also emphasize that within each cycle, the AGN power also varies on very short timescales, which is found in other similar numerical studies as well\citep{Gaspari2012, Prasad2015, Meece2016}. Therefore, the instantaneous radio power of the AGN may not be a good indicator of the feedback strength. 

The success of momentum-driven AGN feedback model in quenching cooling flows in galaxy cluster may not come as a surprise. After all, the cluster potential is rather deep, and even if the coupling of the jet energy to the ICM is not perfect (Figure~\ref{fig:All_average_r}), the self-regulating nature of AGN feedback allows the SMBH to inject enough energy to offset both the radiative cooling loss and the loss of energy outside of the cool core. To further test the model, simulations that follow the metallicity distribution self consistently may be needed \citep{Skory, Meece2016}. More constraints can be put on the model by studying smaller systems with shallower potential \citep{GaspariElliptical, Randall2015}. In addition, we will create mock X-ray observations of the simulated cluster and carry out a systematic comparison between simulations and observations in terms of the scale and nature of the fluctuations.

\section{Conclusions}
\label{conclusions}

We analyze how AGN deposits its energy to the ICM in a simulation where momentum-driven AGN feedback dynamically balances radiative cooling over 6.5 Gyr in an idealized cool-core cluster. We compute the dissipative heating rate due to the artificial viscosity in the Zeus hydro method, and separate shock heating from turbulent heating by identifying shock waves. We analyze how radiative cooling, dissipative heating and heating/cooling due to adiabatic processes change with time and radius. We compare the amount of cooling and heating at different radii to understand the balance between cooling and heating. Our main conclusions are:

1. Within the inner core region ($r<100$ kpc), on average, energy dissipation via shock waves is almost an order of magnitude higher than that via turbulence in the simulation. Most of the shock waves are weak shocks with typical Mach numbers $<1.2$. Thus the velocities of the ICM in the simulation are in concordance with the Hitomi observations of Perseus. Shock dissipation is a steep function of radius, with most of the energy dissipated within $r<30$ kpc, while turbulent dissipation shows a more gradual decline with radius. Radial transportation of heat (radial mixing and adiabatic processes) is important as the post shock materials (hot bubbles) rise and distribute the thermal energy throughout the core. 

2. Dissipative heating itself does not perfectly balance radiative cooling loss spatially or temporally. Some of the jet energy is dissipated outside the cool core, and a considerable fraction (more than half) of the dissipated energy is transported beyond $r>100$ kpc on average, while less than half of the thermal energy is actually used to offset radiative cooling loss within the core. As the cluster experiences cycles of AGN outbursts regulated by the interplay between ICM cooling, star formation and the AGN itself on Gyr timescales, the cluster core undergoes phases of temporary over heating and over cooling. 

3. During the over heating period, sometimes the amount of turbulent dissipation may appear to balance radiative cooling perfectly. However, at the same time, more heat is dissipated vis shock waves and being radially mixed. Thus turbulent heating alone can be enough, but when it is enough, shock dissipation is unavoidably over heating the cluster core.

4. Different interpretations of the X-ray observations of Perseus have resulted in different claims regarding what is the heating mechanism. We find that the Perseus cluster has the characteristics of the over heating phase in our model, with extended multiphase gas and enhanced AGN and star formation activities. Therefore, shock dissipation, turbulent dissipation and bubble mixing can all balance cooling, and are all operating at the same time, temporarily over heating the core of Perseus.

\acknowledgments

We thank Irina Zhuravleva, Peng Oh, Karen Yang, Erwin Lau, Daisuke Nagai and Mark Voit for useful discussions. The authors acknowledge financial support from NASA grants NNX12AH41G and NNX15AB20G, and NSF grants AST-1312888, AST-1615955, AST-0908390, AST-1008134, and AST-1210890. We also acknowledge computing support from Columbia University's Yeti Cluster, as well as computational resources from NSF XSEDE and the University of Michigan. Computations described in this work were performed using the publicly-available Enzo code, which is the product of a collaborative effort of many independent scientists from numerous institutions around the world. Their commitment to open science has helped make this work possible. The simulation data is analyzed using the publicly available yt visualization package \citep{yt}. We are grateful to the yt development team and the yt community for their support. MR is grateful for the hospitality of the UW Astronomy Department, which was made possible in part by a generous gift from Jeff and Joulie Diermeier. 

\appendix
\section{Appendix}
We analyze a clean, simple simulation to test the method that is used to calculate the dissipative heating rate. The simulation is based on the standard test Driven Turbulence 3D which comes with the standard Enzo2.5 (\url{http://enzo-project.org}). We use all the standard parameters except that we change the hydro method to Zeus method, which automatically turns off driving. Thus we are only simulating the decay of supersonic turbulence (with a $k^{-4}$ spectrum). The simulation box has a size of $(10 \rm pc)^3$ with 32 fixed grids (no AMR) in each dimension and periodic boundaries. The gas initially has a uniform density of 100 cm$^{-3}$. There is no gravity or radiative cooling in the simulation. As the turbulence decays, the kinetic energy dissipates into thermal energy. 

\begin{figure}
\begin{center}
\includegraphics[scale=.5]{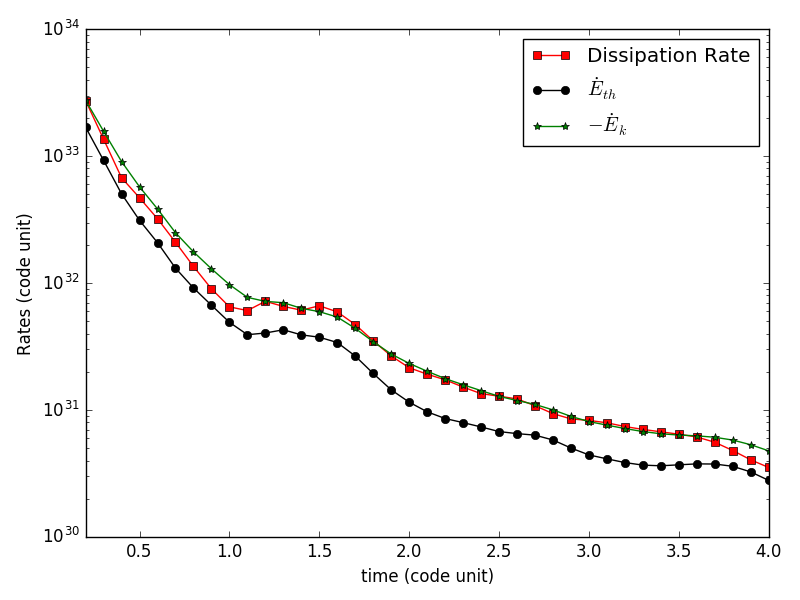} 
\caption{The computed dissipative heating rate (red) compared with the rates at which the kinetic energy decreases (green) and the thermal energy increases (black) measured from the simulation output files. 
\label{fig:paper_test}}
\end{center}
\end{figure}

Figure~\ref{fig:paper_test} shows the dissipative heating rate computed using the method described in Section~\ref{sec:method_2} in red. The black line shows the rate at which the total thermal energy in the box increases, computed as the change in the total thermal energy divided by the time interval $\dot E_{th}= \Delta E_{th} / \Delta t$. The green line shows the rate at which the total kinetic energy in the box decreases, computed in the same fashion. All three lines are in general good agreement with each other, indicating that the calculation does capture the amount of kinetic energy that is dissipated into heat. However, the black line has a lower value than the green line. This is because energy is not conserved when using Zeus method. A fraction (less than 30\%) of the kinetic energy that is dissipated is lost instead of being added to the thermal energy. 

The energy loss is less severe with lower Mach numbers and higher resolution. In the simulation with Mach number of 10, 1.9\% of the total energy is lost at the end of the simulation. When we use an initial Mach number of 2, only less than 0.1\% is lost. With Mach number of 10, when we increase the resolution from $32^3$ to $64^3$ and $256^3$ grids, the energy loss decreases to 1.6\% and 1.4\%. We discuss the consequence of the non-conservation of energy in Section~\ref{sec:discussion_3}.

\begin{thebibliography}{}
\expandafter\ifx\csname natexlab\endcsname\relax\def\natexlab#1{#1}\fi

\bibitem[{{Blanton} {et~al.}(2011){Blanton}, {Randall}, {Clarke}, {Sarazin},
  {McNamara}, {Douglass}, \& {McDonald}}]{Blanton2011}
{Blanton}, E.~L., {Randall}, S.~W., {Clarke}, T.~E., {et~al.} 2011, \apj, 737,
  99

\bibitem[{{Bryan} {et~al.}(2014){Bryan}, {Norman}, {O'Shea}, {Abel}, {Wise},
  {Turk}, {Reynolds}, {Collins}, {Wang}, {Skillman}, {Smith}, {Harkness},
  {Bordner}, {Kim}, {Kuhlen}, {Xu}, {Goldbaum}, {Hummels}, {Kritsuk}, {Tasker},
  {Skory}, {Simpson}, {Hahn}, {Oishi}, {So}, {Zhao}, {Cen}, {Li}, \& {Enzo
  Collaboration}}]{Enzo}
{Bryan}, G.~L., {Norman}, M.~L., {O'Shea}, B.~W., {et~al.} 2014, \apjs, 211, 19

\bibitem[{{Cen} \& {Ostriker}(1992)}]{CenOstriker}
{Cen}, R., \& {Ostriker}, J.~P. 1992, \apjl, 399, L113

\bibitem[{{Churazov} {et~al.}(2001){Churazov}, {Br{\"u}ggen}, {Kaiser},
  {B{\"o}hringer}, \& {Forman}}]{Churazov2001}
{Churazov}, E., {Br{\"u}ggen}, M., {Kaiser}, C.~R., {B{\"o}hringer}, H., \&
  {Forman}, W. 2001, \apj, 554, 261

\bibitem[{{Churazov} {et~al.}(2004){Churazov}, {Forman}, {Jones}, {Sunyaev}, \&
  {B{\"o}hringer}}]{Churazov}
{Churazov}, E., {Forman}, W., {Jones}, C., {Sunyaev}, R., \& {B{\"o}hringer},
  H. 2004, \mnras, 347, 29

\bibitem[{{Conselice} {et~al.}(2001){Conselice}, {Gallagher}, \&
  {Wyse}}]{Conselice2001}
{Conselice}, C.~J., {Gallagher}, III, J.~S., \& {Wyse}, R.~F.~G. 2001, \aj,
  122, 2281

\bibitem[{{Donahue} {et~al.}(2015){Donahue}, {Connor}, {Fogarty}, {Li}, {Voit},
  {Postman}, {Koekemoer}, {Moustakas}, {Bradley}, \& {Ford}}]{Donahue2015}
{Donahue}, M., {Connor}, T., {Fogarty}, K., {et~al.} 2015, \apj, 805, 177

\bibitem[{{Dubois} {et~al.}(2010){Dubois}, {Devriendt}, {Slyz}, \&
  {Teyssier}}]{Dubois2010}
{Dubois}, Y., {Devriendt}, J., {Slyz}, A., \& {Teyssier}, R. 2010, \mnras, 409,
  985

\bibitem[{{Dunn} \& {Fabian}(2006)}]{Dunn2006}
{Dunn}, R.~J.~H., \& {Fabian}, A.~C. 2006, \mnras, 373, 959

\bibitem[{{En{\ss}lin} {et~al.}(2011){En{\ss}lin}, {Pfrommer}, {Miniati}, \&
  {Subramanian}}]{Enblin2011}
{En{\ss}lin}, T., {Pfrommer}, C., {Miniati}, F., \& {Subramanian}, K. 2011,
  \aap, 527, A99

\bibitem[{{Fabian}(1994)}]{Fabian1994}
{Fabian}, A.~C. 1994, \araa, 32, 277

\bibitem[{{Fabian}(2012)}]{Fabian2012}
---. 2012, \araa, 50, 455

\bibitem[{{Fabian} \& {Nulsen}(1977)}]{Fabian1977}
{Fabian}, A.~C., \& {Nulsen}, P.~E.~J. 1977, \mnras, 180, 479

\bibitem[{{Fabian} {et~al.}(2005){Fabian}, {Reynolds}, {Taylor}, \&
  {Dunn}}]{Fabian2005}
{Fabian}, A.~C., {Reynolds}, C.~S., {Taylor}, G.~B., \& {Dunn}, R.~J.~H. 2005,
  \mnras, 363, 891

\bibitem[{{Fabian} {et~al.}(2003){Fabian}, {Sanders}, {Allen}, {Crawford},
  {Iwasawa}, {Johnstone}, {Schmidt}, \& {Taylor}}]{Fabian2003}
{Fabian}, A.~C., {Sanders}, J.~S., {Allen}, S.~W., {et~al.} 2003, \mnras, 344,
  L43

\bibitem[{{Fabian} {et~al.}(2006){Fabian}, {Sanders}, {Taylor}, {Allen},
  {Crawford}, {Johnstone}, \& {Iwasawa}}]{Fabian2006}
{Fabian}, A.~C., {Sanders}, J.~S., {Taylor}, G.~B., {et~al.} 2006, \mnras, 366,
  417

\bibitem[{{Gaspari} {et~al.}(2012{\natexlab{a}}){Gaspari}, {Brighenti}, \&
  {Temi}}]{GaspariElliptical}
{Gaspari}, M., {Brighenti}, F., \& {Temi}, P. 2012{\natexlab{a}}, \mnras, 424,
  190

\bibitem[{{Gaspari} {et~al.}(2011){Gaspari}, {Melioli}, {Brighenti}, \&
  {D'Ercole}}]{Gaspari2011}
{Gaspari}, M., {Melioli}, C., {Brighenti}, F., \& {D'Ercole}, A. 2011, \mnras,
  411, 349

\bibitem[{{Gaspari} {et~al.}(2012{\natexlab{b}}){Gaspari}, {Ruszkowski}, \&
  {Sharma}}]{Gaspari2012}
{Gaspari}, M., {Ruszkowski}, M., \& {Sharma}, P. 2012{\natexlab{b}}, \apj, 746,
  94

\bibitem[{{Guo} \& {Oh}(2008)}]{GO2008}
{Guo}, F., \& {Oh}, S.~P. 2008, \mnras, 384, 251

\bibitem[{{Hillel} \& {Soker}(2016)}]{Hillel2016}
{Hillel}, S., \& {Soker}, N. 2016, \mnras, 455, 2139

\bibitem[{{Hitomi Collaboration} {et~al.}(2016){Hitomi Collaboration},
  {Aharonian}, {Akamatsu}, {Akimoto}, {Allen}, {Anabuki}, {Angelini}, {Arnaud},
  {Audard}, {Awaki}, {Axelsson}, {Bamba}, {Bautz}, {Blandford}, {Brenneman},
  {Brown}, {Bulbul}, {Cackett}, {Chernyakova}, {Chiao}, {Coppi}, {Costantini},
  {de Plaa}, {den Herder}, {Done}, {Dotani}, {Ebisawa}, {Eckart}, {Enoto},
  {Ezoe}, {Fabian}, {Ferrigno}, {Foster}, {Fujimoto}, {Fukazawa}, {Furuzawa},
  {Galeazzi}, {Gallo}, {Gandhi}, {Giustini}, {Goldwurm}, {Gu}, {Guainazzi},
  {Haba}, {Hagino}, {Hamaguchi}, {Harrus}, {Hatsukade}, {Hayashi}, {Hayashi},
  {Hayashida}, {Hiraga}, {Hornschemeier}, {Hoshino}, {Hughes}, {Iizuka},
  {Inoue}, {Inoue}, {Ishibashi}, {Ishida}, {Ishikawa}, {Ishisaki}, {Itoh},
  {Iyomoto}, {Kaastra}, {Kallman}, {Kamae}, {Kara}, {Kataoka}, {Katsuda},
  {Katsuta}, {Kawaharada}, {Kawai}, {Kelley}, {Khangulyan}, {Kilbourne},
  {King}, {Kitaguchi}, {Kitamoto}, {Kitayama}, {Kohmura}, {Kokubun}, {Koyama},
  {Koyama}, {Kretschmar}, {Krimm}, {Kubota}, {Kunieda}, {Laurent}, {Lebrun},
  {Lee}, {Leutenegger}, {Limousin}, {Loewenstein}, {Long}, {Lumb}, {Madejski},
  {Maeda}, {Maier}, {Makishima}, {Markevitch}, {Matsumoto}, {Matsushita},
  {McCammon}, {McNamara}, {Mehdipour}, {Miller}, {Miller}, {Mineshige},
  {Mitsuda}, {Mitsuishi}, {Miyazawa}, {Mizuno}, {Mori}, {Mori}, {Moseley},
  {Mukai}, {Murakami}, {Murakami}, {Mushotzky}, {Nagino}, {Nakagawa},
  {Nakajima}, {Nakamori}, {Nakano}, {Nakashima}, {Nakazawa}, {Nobukawa},
  {Noda}, {Nomachi}, {O'Dell}, {Odaka}, {Ohashi}, {Ohno}, {Okajima}, {Ota},
  {Ozaki}, {Paerels}, {Paltani}, {Parmar}, {Petre}, {Pinto}, {Pohl}, {Porter},
  {Pottschmidt}, {Ramsey}, {Reynolds}, {Russell}, {Safi-Harb}, {Saito},
  {Sakai}, {Sameshima}, {Sato}, {Sato}, {Sato}, {Sawada}, {Schartel},
  {Serlemitsos}, {Seta}, {Shidatsu}, {Simionescu}, {Smith}, {Soong}, {Stawarz},
  {Sugawara}, {Sugita}, {Szymkowiak}, {Tajima}, {Takahashi}, {Takahashi},
  {Takeda}, {Takei}, {Tamagawa}, {Tamura}, {Tamura}, {Tanaka}, {Tanaka},
  {Tanaka}, {Tashiro}, {Tawara}, {Terada}, {Terashima}, {Tombesi}, {Tomida},
  {Tsuboi}, {Tsujimoto}, {Tsunemi}, {Tsuru}, {Uchida}, {Uchiyama}, {Uchiyama},
  {Ueda}, {Ueda}, {Ueno}, {Uno}, {Urry}, {Ursino}, {de Vries}, {Watanabe},
  {Werner}, {Wik}, {Wilkins}, {Williams}, {Yamada}, {Yamaguchi}, {Yamaoka},
  {Yamasaki}, {Yamauchi}, {Yamauchi}, {Yaqoob}, {Yatsu}, {Yonetoku}, {Yoshida},
  {Yuasa}, {Zhuravleva}, \& {Zoghbi}}]{Hitomi2016}
{Hitomi Collaboration}, {Aharonian}, F., {Akamatsu}, H., {et~al.} 2016, \nat,
  535, 117

\bibitem[{{Hlavacek-Larrondo} {et~al.}(2012){Hlavacek-Larrondo}, {Fabian},
  {Edge}, {Ebeling}, {Sanders}, {Hogan}, \& {Taylor}}]{Hlavacek2012}
{Hlavacek-Larrondo}, J., {Fabian}, A.~C., {Edge}, A.~C., {et~al.} 2012, \mnras,
  421, 1360

\bibitem[{{Hoffer} {et~al.}(2012){Hoffer}, {Donahue}, {Hicks}, \&
  {Barthelemy}}]{Hoffer2012}
{Hoffer}, A.~S., {Donahue}, M., {Hicks}, A., \& {Barthelemy}, R.~S. 2012,
  \apjs, 199, 23

\bibitem[{{Kunz} {et~al.}(2011){Kunz}, {Schekochihin}, {Cowley}, {Binney}, \&
  {Sanders}}]{Kunz2011}
{Kunz}, M.~W., {Schekochihin}, A.~A., {Cowley}, S.~C., {Binney}, J.~J., \&
  {Sanders}, J.~S. 2011, \mnras, 410, 2446

\bibitem[{{Li} \& {Bryan}(2014{\natexlab{a}})}]{PIII}
{Li}, Y., \& {Bryan}, G.~L. 2014{\natexlab{a}}, \apj, 789, 54

\bibitem[{{Li} \& {Bryan}(2014{\natexlab{b}})}]{PII}
---. 2014{\natexlab{b}}, \apj, 789, 153

\bibitem[{{Li} {et~al.}(2015){Li}, {Bryan}, {Ruszkowski}, {Voit}, {O'Shea}, \&
  {Donahue}}]{Li2015}
{Li}, Y., {Bryan}, G.~L., {Ruszkowski}, M., {et~al.} 2015, \apj, 811, 73

\bibitem[{{Lyutikov}(2006)}]{Lyutikov2006}
{Lyutikov}, M. 2006, \mnras, 373, 73

\bibitem[{{Mathews} {et~al.}(2003){Mathews}, {Brighenti}, {Buote}, \&
  {Lewis}}]{Mathews2003}
{Mathews}, W.~G., {Brighenti}, F., {Buote}, D.~A., \& {Lewis}, A.~D. 2003,
  \apj, 596, 159

\bibitem[{{Mathews} {et~al.}(2006){Mathews}, {Faltenbacher}, \&
  {Brighenti}}]{Mathews}
{Mathews}, W.~G., {Faltenbacher}, A., \& {Brighenti}, F. 2006, \apj, 638, 659

\bibitem[{{McDonald} {et~al.}(2010){McDonald}, {Veilleux}, {Rupke}, \&
  {Mushotzky}}]{McDonald10}
{McDonald}, M., {Veilleux}, S., {Rupke}, D.~S.~N., \& {Mushotzky}, R. 2010,
  \apj, 721, 1262

\bibitem[{{McNamara} \& {Nulsen}(2007)}]{McNamara2007}
{McNamara}, B.~R., \& {Nulsen}, P.~E.~J. 2007, \araa, 45, 117

\bibitem[{{McNamara} {et~al.}(2005){McNamara}, {Nulsen}, {Wise}, {Rafferty},
  {Carilli}, {Sarazin}, \& {Blanton}}]{McNamara2005}
{McNamara}, B.~R., {Nulsen}, P.~E.~J., {Wise}, M.~W., {et~al.} 2005, \nat, 433,
  45

\bibitem[{{McNamara} \& {O'Connell}(1989)}]{McNamara1989}
{McNamara}, B.~R., \& {O'Connell}, R.~W. 1989, \aj, 98, 2018

\bibitem[{{Meece} {et~al.}(2016){Meece}, {Voit}, \& {O'Shea}}]{Meece2016}
{Meece}, G.~R., {Voit}, G.~M., \& {O'Shea}, B.~W. 2016, ArXiv e-prints,
  arXiv:1603.03674

\bibitem[{{Merloni} \& {Heinz}(2007)}]{Merloni2007}
{Merloni}, A., \& {Heinz}, S. 2007, \mnras, 381, 589

\bibitem[{{Mittal} {et~al.}(2012){Mittal}, {Oonk}, {Ferland}, {Edge}, {O'Dea},
  {Baum}, {Whelan}, {Johnstone}, {Combes}, {Salom{\'e}}, {Fabian}, {Tremblay},
  {Donahue}, \& {Russell}}]{Mittal2012}
{Mittal}, R., {Oonk}, J.~B.~R., {Ferland}, G.~J., {et~al.} 2012, \mnras, 426,
  2957

\bibitem[{{Navarro} {et~al.}(1996){Navarro}, {Frenk}, \& {White}}]{NFW}
{Navarro}, J.~F., {Frenk}, C.~S., \& {White}, S.~D.~M. 1996, \apj, 462, 563

\bibitem[{{Nulsen} {et~al.}(2005){Nulsen}, {McNamara}, {Wise}, \&
  {David}}]{Nulsen2005}
{Nulsen}, P.~E.~J., {McNamara}, B.~R., {Wise}, M.~W., \& {David}, L.~P. 2005,
  \apj, 628, 629

\bibitem[{{O'Dea} {et~al.}(2008){O'Dea}, {Baum}, {Privon}, {Noel-Storr},
  {Quillen}, {Zufelt}, {Park}, {Edge}, {Russell}, {Fabian}, {Donahue},
  {Sarazin}, {McNamara}, {Bregman}, \& {Egami}}]{ODea2008}
{O'Dea}, C.~P., {Baum}, S.~A., {Privon}, G., {et~al.} 2008, \apj, 681, 1035

\bibitem[{{Omma} {et~al.}(2004){Omma}, {Binney}, {Bryan}, \& {Slyz}}]{Omma2004}
{Omma}, H., {Binney}, J., {Bryan}, G., \& {Slyz}, A. 2004, \mnras, 348, 1105

\bibitem[{{Paterno-Mahler} {et~al.}(2013){Paterno-Mahler}, {Blanton},
  {Randall}, \& {Clarke}}]{Rachel2013}
{Paterno-Mahler}, R., {Blanton}, E.~L., {Randall}, S.~W., \& {Clarke}, T.~E.
  2013, \apj, 773, 114

\bibitem[{{Peterson} {et~al.}(2003){Peterson}, {Kahn}, {Paerels}, {Kaastra},
  {Tamura}, {Bleeker}, {Ferrigno}, \& {Jernigan}}]{Peterson2003}
{Peterson}, J.~R., {Kahn}, S.~M., {Paerels}, F.~B.~S., {et~al.} 2003, \apj,
  590, 207

\bibitem[{{Prasad} {et~al.}(2015){Prasad}, {Sharma}, \& {Babul}}]{Prasad2015}
{Prasad}, D., {Sharma}, P., \& {Babul}, A. 2015, \apj, 811, 108

\bibitem[{{Randall} {et~al.}(2015){Randall}, {Nulsen}, {Jones}, {Forman},
  {Bulbul}, {Clarke}, {Kraft}, {Blanton}, {David}, {Werner}, {Sun}, {Donahue},
  {Giacintucci}, \& {Simionescu}}]{Randall2015}
{Randall}, S.~W., {Nulsen}, P.~E.~J., {Jones}, C., {et~al.} 2015, \apj, 805,
  112

\bibitem[{{Reynolds} {et~al.}(2015){Reynolds}, {Balbus}, \&
  {Schekochihin}}]{Reynolds2015}
{Reynolds}, C.~S., {Balbus}, S.~A., \& {Schekochihin}, A.~A. 2015, \apj, 815,
  41

\bibitem[{{Reynolds} {et~al.}(2005){Reynolds}, {McKernan}, {Fabian}, {Stone},
  \& {Vernaleo}}]{Reynolds2005}
{Reynolds}, C.~S., {McKernan}, B., {Fabian}, A.~C., {Stone}, J.~M., \&
  {Vernaleo}, J.~C. 2005, \mnras, 357, 242

\bibitem[{{Ruszkowski} {et~al.}(2004){Ruszkowski}, {Br{\"u}ggen}, \&
  {Begelman}}]{Ruszkowski2004}
{Ruszkowski}, M., {Br{\"u}ggen}, M., \& {Begelman}, M.~C. 2004, \apj, 611, 158

\bibitem[{{Ruszkowski} {et~al.}(2008){Ruszkowski}, {En{\ss}lin}, {Br{\"u}ggen},
  {Begelman}, \& {Churazov}}]{Ruszkowski2008}
{Ruszkowski}, M., {En{\ss}lin}, T.~A., {Br{\"u}ggen}, M., {Begelman}, M.~C., \&
  {Churazov}, E. 2008, \mnras, 383, 1359

\bibitem[{{Ryu} {et~al.}(2003){Ryu}, {Kang}, {Hallman}, \& {Jones}}]{Ryu2003}
{Ryu}, D., {Kang}, H., {Hallman}, E., \& {Jones}, T.~W. 2003, \apj, 593, 599

\bibitem[{{Schmidt} {et~al.}(2002){Schmidt}, {Fabian}, \&
  {Sanders}}]{Metallicity}
{Schmidt}, R.~W., {Fabian}, A.~C., \& {Sanders}, J.~S. 2002, \mnras, 337, 71

\bibitem[{{Skillman} {et~al.}(2011){Skillman}, {Hallman}, {O'Shea}, {Burns},
  {Smith}, \& {Turk}}]{Skillman2011}
{Skillman}, S.~W., {Hallman}, E.~J., {O'Shea}, B.~W., {et~al.} 2011, \apj, 735,
  96

\bibitem[{{Skory} {et~al.}(2013){Skory}, {Hallman}, {Burns}, {Skillman},
  {O'Shea}, \& {Smith}}]{Skory}
{Skory}, S., {Hallman}, E., {Burns}, J.~O., {et~al.} 2013, \apj, 763, 38

\bibitem[{{Sternberg} \& {Soker}(2009)}]{Sternberg2009}
{Sternberg}, A., \& {Soker}, N. 2009, \mnras, 395, 228

\bibitem[{{Stone} \& {Norman}(1992{\natexlab{a}})}]{Zeus}
{Stone}, J.~M., \& {Norman}, M.~L. 1992{\natexlab{a}}, \apjs, 80, 753

\bibitem[{{Stone} \& {Norman}(1992{\natexlab{b}})}]{Zeus2}
---. 1992{\natexlab{b}}, \apjs, 80, 791

\bibitem[{{Tasker} \& {Bryan}(2006)}]{Tasker2006}
{Tasker}, E.~J., \& {Bryan}, G.~L. 2006, \apj, 641, 878

\bibitem[{{Tremblay} {et~al.}(2015){Tremblay}, {O'Dea}, {Baum}, {Mittal},
  {McDonald}, {Combes}, {Li}, {McNamara}, {Bremer}, {Clarke}, {Donahue},
  {Edge}, {Fabian}, {Hamer}, {Hogan}, {Oonk}, {Quillen}, {Sanders},
  {Salom{\'e}}, \& {Voit}}]{Tremblay2015}
{Tremblay}, G.~R., {O'Dea}, C.~P., {Baum}, S.~A., {et~al.} 2015, \mnras, 451,
  3768

\bibitem[{{Turk} {et~al.}(2011){Turk}, {Smith}, {Oishi}, {Skory}, {Skillman},
  {Abel}, \& {Norman}}]{yt}
{Turk}, M.~J., {Smith}, B.~D., {Oishi}, J.~S., {et~al.} 2011, \apjs, 192, 9

\bibitem[{{Vernaleo} \& {Reynolds}(2006)}]{Reynolds06}
{Vernaleo}, J.~C., \& {Reynolds}, C.~S. 2006, \apj, 645, 83

\bibitem[{{Voit} \& {Donahue}(2005)}]{VoitDonahue2005}
{Voit}, G.~M., \& {Donahue}, M. 2005, \apj, 634, 955

\bibitem[{{Voit} {et~al.}(2016){Voit}, {Meece}, {Li}, {O'Shea}, {Bryan}, \&
  {Donahue}}]{Mark2016}
{Voit}, G.~M., {Meece}, G., {Li}, Y., {et~al.} 2016, ArXiv e-prints,
  arXiv:1607.02212

\bibitem[{{Wilman} {et~al.}(2005){Wilman}, {Edge}, \& {Johnstone}}]{BHmass}
{Wilman}, R.~J., {Edge}, A.~C., \& {Johnstone}, R.~M. 2005, \mnras, 359, 755

\bibitem[{{Wise} {et~al.}(2007){Wise}, {McNamara}, {Nulsen}, {Houck}, \&
  {David}}]{Wise2007}
{Wise}, M.~W., {McNamara}, B.~R., {Nulsen}, P.~E.~J., {Houck}, J.~C., \&
  {David}, L.~P. 2007, \apj, 659, 1153

\bibitem[{{Yang} \& {Reynolds}(2016)}]{Yang2016}
{Yang}, H.-Y.~K., \& {Reynolds}, C.~S. 2016, ArXiv e-prints, arXiv:1605.01725

\bibitem[{{Zhuravleva} {et~al.}(2014){Zhuravleva}, {Churazov}, {Schekochihin},
  {Allen}, {Ar{\'e}valo}, {Fabian}, {Forman}, {Sanders}, {Simionescu},
  {Sunyaev}, {Vikhlinin}, \& {Werner}}]{Zhuravleva2014}
{Zhuravleva}, I., {Churazov}, E., {Schekochihin}, A.~A., {et~al.} 2014, \nat,
  515, 85

\bibitem[{{Zhuravleva} {et~al.}(2016){Zhuravleva}, {Churazov}, {Ar{\'e}valo},
  {Schekochihin}, {Forman}, {Allen}, {Simionescu}, {Sunyaev}, {Vikhlinin}, \&
  {Werner}}]{Zhuravleva2016}
{Zhuravleva}, I., {Churazov}, E., {Ar{\'e}valo}, P., {et~al.} 2016, \mnras,
  458, 2902

\bibitem[{{ZuHone} {et~al.}(2010){ZuHone}, {Markevitch}, \&
  {Johnson}}]{ZuHone2010}
{ZuHone}, J.~A., {Markevitch}, M., \& {Johnson}, R.~E. 2010, \apj, 717, 908

\end{thebibliography}
\end{document}